\RequirePackage{lineno}
\documentclass[aps,prl,twocolumn,superscriptaddress,groupedaddress]{revtex4}  % for review and submission

\usepackage{graphicx}  % needed for figures
\usepackage{dcolumn}   % needed for some tables
\usepackage{bm}        % for math
\usepackage{amssymb}   % for math
\usepackage{epsfig}
\usepackage{epstopdf}
\usepackage{amsmath}
\usepackage[latin9]{inputenc}

\begin{document}

%\begin{linenumbers}
% The following information is for internal review, please remove them for submission
%\widetext
%\leftline{Version xx as of \today}
%\leftline{Primary authors: Joe E. Physics}
%\leftline{To be submitted to (PRL, PRD-RC, PRD, PLB; choose one.)}
%\leftline{Comment to {\tt d0-run2eb-nnn@fnal.gov} by xxx, yyy}
%\centerline{\em D\O\ INTERNAL DOCUMENT -- NOT FOR PUBLIC DISTRIBUTION}

% the following line is for submission, including submission to the arXiv!!
%\hspace{5.2in} \mbox{Fermilab-Pub-04/xxx-E}

\title{A laser excitation scheme for $^{229\mathrm{m}}$Th}
\author{Lars von der Wense}
\email{L.Wense@physik.uni-muenchen.de}
\author{Benedict Seiferle}
\affiliation{Ludwig-Maximilians-Universität München, 85748 Garching, Germany.}
\author{Simon Stellmer}
\affiliation{Technische Universität Wien, 1040 Vienna, Austria.}
\author{Johannes Weitenberg}
\affiliation{Max-Planck-Institut für Quantenoptik, 85748 Garching, Germany.}
\author{Georgy Kazakov}
\affiliation{Technische Universität Wien, 1040 Vienna, Austria.}
\author{Adriana Pálffy}
\affiliation{Max-Planck-Institut für Kernphysik, 69117 Heidelberg, Germany.}
\author{Peter G. Thirolf}
\affiliation{Ludwig-Maximilians-Universität München, 85748 Garching, Germany.}
\date{\today}
\begin{abstract}
%\linenumbers
\noindent Direct laser excitation of the lowest known nuclear excited state in $^{229}$Th has been a longstanding objective. It is generally assumed that reaching this goal would require a considerably reduced uncertainty of the isomer's excitation energy compared to the presently adopted value of $(7.8\pm0.5)$~eV. Here we present a direct laser excitation scheme for $^{229\mathrm{m}}$Th, which circumvents this requirement. The proposed excitation scheme makes use of already existing laser technology and therefore paves the way for nuclear laser spectroscopy. In this concept, the recently experimentally observed internal-conversion decay channel of the isomeric state is used for probing the isomeric population. A signal-to-background ratio of better than $10^4$ and a total measurement time of less than three days for laser scanning appear to be achievable.
\end{abstract}
\maketitle

\noindent Direct nuclear laser excitation has come closer into reach within the past years, partly due to the development of free-electron-laser technology. Such ideas are typically dealing with the excitation of nuclear states in the energy range of at least a few keV or above \cite{Evers,Palffy}. However, there is one exceptional nuclear state known for the last 40 years with a significantly lower energy of presumably below 10~eV \cite{Kroger_Reich,Helmer_Reich2}. An excitation energy of $(7.8\pm0.5)$~eV, corresponding to $(159\pm11)$~nm wavelength or $\sim1900$~THz, is by today the most accepted value for the isomeric first excited state of $^{229}$Th \cite{Beck1,Beck2}. This state conceptionally allows for direct nuclear laser excitation using solid-state laser technology and was proposed for the development of a nuclear clock of extremely high stability, due to an expected high resilience against external influences and a radiative lifetime in the range of minutes to hours \cite{Tkalya1,Peik1,Campbell2,Rellergert}.\\
It is generally assumed that direct nuclear laser excitation of $^{229\mathrm{m}}$Th requires a considerably improved knowledge of the isomeric transition energy (see e.g. Ref.~\cite{Peik3} and references therein).
The reasons are that, first, the present uncertainty in the knowledge of the isomeric energy value is still rather large. The currently best energy value of $7.8\pm0.5$~eV leads to an energy range of at least 1~eV (corresponding to $2.4\cdot10^{14}$~Hz) to be scanned when searching for the isomeric excitation. Second, the radiative lifetime of $^{229\text{m}}$Th was theoretically predicted to be in the range of hours \cite{Ruchowska,Tkalya4,Minkov}, leading to long required detection times when searching for a radiative decay channel. This has led to the assumption that the required time for laser-based scanning of the large energy range of 1~eV would be prohibitively long.\\
In order to shorten the required scanning times, there are worldwide efforts ongoing to decrease the uncertainty of the transition energy, which would bring the isomeric state into realistic reach of direct laser excitation (see e.g. Refs.~\cite{Seiferle4,Kazakov,Yamaguchi,Jeet}). A recent review is found in Ref.~\cite{Peik3}.\\
Here we propose a different approach, which allows for a direct laser excitation of $^{229\mathrm{m}}$Th without the requirement of an improved knowledge of the transition energy. As the required laser technology is already available, this proposal paves the way for direct nuclear laser spectroscopy of $^{229\text{m}}$Th.\\
As opposed to previous experiments, this idea is making use of the fast ($\sim10\ \mu$s lifetime) non-radiative internal conversion (IC) decay channel of neutral $^{229\mathrm{m}}$Th for the isomer detection. During IC decay, the nuclear excitation energy is transferred to the atomic shell, leading to the ejection of a shell electron. The IC decay channel of $^{229\text{m}}$Th has recently been experimentally observed \cite{Wense1,Seiferle3} and is known to be the dominant decay channel under the considered experimental conditions. This approach corresponds to laser-based conversion electron M\"ossbauer spectroscopy (CEMS) in the optical region \cite{Salvat} and is found to be advantageous compared to earlier proposals that make use of a potential radiative decay for isomer detection. The reason for this improved perspective is the $\sim9$ orders of magnitude faster isomeric IC decay, which allows to trigger the decay detection correlated with the laser pulses and in this way to significantly improve the signal-to-background ratio of the detection, while shortening the required time to search for the direct nuclear laser excitation of $^{229\mathrm{m}}$Th to about three days, as will be detailed in the following.\\
\noindent The on-resonance laser irradiation of a two-level system of a nuclear ground and excited state can be modeled in case of low laser intensity and a laser bandwidth significantly broader than the nuclear transition linewidth via the Einstein rate equations \cite{Loudon,Hilborn}

\begin{equation}
\begin{aligned}
\dot{N}_\text{gnd}&=+\Gamma_\text{tot} N_\text{exc}+\rho^\omega B^\omega N_\text{exc}-\rho^\omega B^\omega \frac{g_\text{exc}}{g_\text{gnd}}N_\text{gnd}\\
\dot{N}_\text{exc}&=-\Gamma_\text{tot} N_\text{exc}-\rho^\omega B^\omega N_\text{exc}+\rho^\omega B^\omega \frac{g_\text{exc}}{g_\text{gnd}}N_\text{gnd}.
\end{aligned}
\label{einstein}
\end{equation}
Here $N_\text{gnd}$ and $N_\text{exc}$ denote the number of nuclei in the ground and excited state, respectively, $\rho^{\omega}$ is the spectral energy density of the electromagnetic field which drives the transition and $\Gamma_{\text{tot}}=\left(1+\alpha_{\text{ic}}\right)A$ is the total transition rate of the nuclear excitation including both, radiative decay with rate $A$ and IC. $\alpha_{\text{ic}}$ denotes the IC coefficient of the nuclear excited state, defined as the fraction of IC compared to radiative decay. $g_\text{gnd}=2 j_\text{gnd}+1$ and $g_\text{exc}=2 j_\text{exc}+1$ are the degeneracies of the ground and excited nuclear levels, which are calculated from the angular momentum quantum numbers $j_\text{gnd}=5/2$ and $j_\text{exc}=3/2$ to be $6$ and $4$ for the ground and excited state of $^{229}$Th, respectively. $B^\omega$ is the Einstein B-coefficient, which is related to $A$ via \cite{Hilborn} $B^\omega=\left(\pi^2 c^3 A\right)/\left(\hbar\omega^3\right)$, with $\omega$ the angular frequency corresponding to the nuclear transition (taking a value of $\omega\approx 2\pi\cdot1900$~THz in case of $^{229\text{m}}$Th), $c$ the speed of light and $\hbar$ the reduced Planck constant.\\
Solving the Einstein rate equations Eq.~(\ref{einstein}) for the initial conditions $N_\text{gnd}(0)=N_{0}$ (where $N_{0}$ denotes the total number of irradiated nuclei) and $N_\text{exc}(0)=0$, leads to the number of excited nuclei in dependence on the excitation time $t_\text{exc}$ to be \cite{Wense3}

\begin{equation}
N_\text{exc}(t_\text{exc})= \frac{\rho^{\omega} N_{0} \pi^2 c^3}{\left(1+\alpha_{\text{ic}}\right)\hbar \omega^3}\frac{g_\text{exc}}{g_\text{gnd}}\left(1-\mathrm{e}^{-\Gamma_{\text{tot}}t_\text{exc}}\right).
\label{laser1}
\end{equation}
Here it was assumed that $\rho^\omega B^\omega\ll \Gamma_\text{tot}$, which corresponds to the assumption of a low laser intensity and will be fulfilled for most experimental conditions.\\
The radiative decay rate $A$ is related to the radiative isomeric lifetime $\tau_\gamma$ via $A=1/\tau_\gamma$. In case of $^{229\text{m}}$Th no conclusive experimental value for the radiative isomeric lifetime has been reported and there is currently no consensus on $\tau_\gamma$ from theory \cite{Ruchowska,Tkalya4,Minkov}. For the following we will conservatively assume a value of $\tau_\gamma\approx10^4$~s, which lies on the upper limit of the theoretically predicted isomeric lifetime and for this reason leads to the smallest coupling. Correspondingly, the radiative decay rate $A$ is inferred to be $A\approx10^{-4}\ \mathrm{s}^{-1}$.\\
The value for $\alpha_{\text{ic}}$ is subject to significant variation, depending on the electronic surrounding of the $^{229}$Th nucleus. When assuming charged $^{229\text{m}}$Th ions in their atomic shell ground states, the internal conversion decay channel should be energetically suppressed, leading to expectedly $\alpha_{\text{ic}}=0$. In the considered case of neutral $^{229}$Th atoms, the internal conversion decay is allowed due to an isomeric energy exceeding the thorium ionization potential of 6.31~eV \cite{Trautmann}. This results in a reduced isomeric lifetime of $\tau_{\text{tot}}=1/\Gamma_{\text{tot}}\approx10$~$\mu$s \cite{Strizhov,Karpeshin1}, which was recently experimentally confirmed \cite{Seiferle3}. From the assumption of a radiative isomeric lifetime of $\tau_\gamma\approx10^4$~s, the internal conversion coefficient can thus be inferred to take a value of $\alpha_{\text{ic}}\approx10^{9}$. This corresponds to a 9 orders of magnitude IC-broadened nuclear transition linewidth of 15.9~kHz, compared to the case when IC is suppressed.\\[0.2cm]
\noindent For the following calculations, a tunable and pulsed VUV laser source with a pulse energy of $E_L=10$~$\mu$J around 160~nm is assumed, providing a bandwidth of $\Delta\nu_L=10$~GHz, a pulse duration of $T_L=5$~ns and a repetition rate of $R_L=10$~Hz, as was already developed in 2009 based on resonance-enhanced four wave mixing \cite{Hanna}. The energy of $10$~$\mu$J per pulse is achieved when the laser is tuned in an energy region between 7.4~eV and 8.4~eV. Up to energies of 9.5~eV a decrease in laser power by a factor of about 100 occurs. It is however emphasized that the expected high signal-to-background ratio of the proposed concept of beyond $10^4$ allows to compensate for this decrease in pulse energy. In case that the isomeric transition is not detected in this energy interval, the search region could be further extended in a second step. Isomeric energies of below 7.4~eV could be covered by a different laser system based on frequency conversion in KBBF crystals \cite{Dai}. The proposed concept can therefore be applied to probe the isomeric excitation in an energy region that covers three standard deviations of the expected energy range of $(7.8\pm0.5)$~eV proposed in Refs.~\cite{Beck1,Beck2}. Energies between 9.5~eV and 11.3~eV could in principle also be covered by applying a tunable VUV laser system that provides a smaller bandwidth in order to compensate for the reduced pulse energy \cite{Hilbig}. In this way the nuclear excitation rate could be increased at the cost of a prolonged measurement time without affecting the experimental concept. However, further difficulties may arise during experiments with radiation in the deep VUV.\\
If the laser light is focused to an area of $A_L=1$~mm$^2$, the spectral energy density $\rho^\omega$ for a single pulse is calculated to be

\begin{equation}
\rho^\omega=\frac{E_L}{cT_LA_L\cdot 2\pi\Delta\nu_L}\approx 1.1\cdot10^{-10}\ \mathrm{Jm}^{-3}\mathrm{Hz}^{-1}.
\end{equation}
Let us now assume an experimental setup where the light irradiates a few nm thin layer of metallic $^{229}$Th deposited onto a gold surface. One might expect that the laser irradiation will lead to a significant ablation of surface atoms. For the proposed experiment the laser irradiance amounts to only $2\cdot10^5$~Wcm$^{-2}$ and is therefore three orders of magnitude below the typical laser ablation threshold for metals \cite{Zimmermann}. Nevertheless a minor desorption of surface atoms may still occur below the ablation threshold \cite{Gill}, leading to a reduction of irradiated thorium atoms after several $10^4$ laser pulses. Such effects could be compensated for by using a moving (e.g. rotating or tape station) target. In this way the number of irradiated thorium atoms would be kept approximately constant.\\
In such an experiment, any isomeric decay would securely occur by internal conversion (IC) under emission of an electron, as the work function of metallic thorium is $3.7$~eV \cite{Riviere} and thus significantly below the expected isomeric energy value. The mean free path length of electrons of 7.8~eV energy above the Fermi level in solids can be estimated to 2.5~nm \cite{Seah}. If the thorium layer is chosen in accordance with this value, about 25\% of the IC electrons would be able to leave the target material. The density of metallic $^{229}$Th is 11.57~gcm$^{-3}$, such that a 2.5~nm thin layer with an area of 1~mm$^{2}$ contains a number of $N_0\approx 7.6\cdot10^{13}$ $^{229}$Th atoms. As $^{229}$Th is an $\alpha$ emitter with 7932 years half-life, this corresponds to an activity of about 210~Bq. Under realistic experimental conditions oxidation might occur. However, the layer thickness as well as the mean free path length of the electrons will not significantly change for $^{229}$ThO$_2$ compared to metallic thorium. Also the work function after oxidation would remain small \cite{Riviere2}. Therefore the isomeric decay would still occur by IC and the experimental concept remains unaffected. Potential surface contaminations due to water or hydrocarbons could be efficiently reduced by providing ultra-high vacuum conditions. The time averaged VUV laser irradiance is $0.1$ mW/mm$^{2}$ and does not lead to a significant carbon-layer growth \cite{Hollenshead}.\\
Experimental data for the optical penetration depth of VUV radiation in metals is scarce. The optical penetration depth in gold was studied and it was found that the extinction coefficient for light with a wavelength around 160~nm is about $k=1.1$ \cite{Canfield}, leading to an optical penetration depth of $\sim11.6$~nm. It is reasonable to assume that the optical penetration depth for thorium in the considered wavelength region is in the same range. For this reason no significant decrease in laser intensity during propagation through the 2.5~nm thin layer of thorium has to be expected. The same holds for the on-resonance absorption of $^{229}$Th. The attenuation coefficient $\mu$ originating from the isomer's absorption is calculated from the absorption cross section $\sigma$ as $\mu=\sigma N$, with $N$ the number of nuclei per volume. The maximum cross section is obtained at the center of the absorption line. For this case one has \cite{Hilborn} $\sigma(\omega)=2\pi g_\text{exc}/g_\text{gnd} (c/\omega)^2 A/\Gamma_\text{tot}\approx 2.7\cdot10^{-18}$~mm$^{2}$. With an assumed number density of $N\approx3\cdot10^{19}$~mm$^{-3}$, the on-resonance attenuation coefficient is calculated to be $\mu(\omega)\approx 81$~mm$^{-1}$, resulting in an optical penetration depth of about 12.3~$\mu$m, significantly larger than the $^{229}$Th layer thickness.\\
Based on Eq.~(\ref{laser1}), the number of excited $^{229\text{m}}$Th nuclei at the end of the laser pulse at $t_\text{exc}=5$~ns is calculated to be $N_\text{exc}(t_\text{exc})\approx 4.2\cdot10^3$. These $^{229\mathrm{m}}$Th nuclei will decay with the IC-enhanced decay rate of $\Gamma_\mathrm{tot}=(1+\alpha_\text{ic})A\approx 10^5$~s$^{-1}$ via electron emission. Therefore, within a decay time of $t_\text{dec}=1/\Gamma_\text{tot}\approx 10$~$\mu$s, the number of isomeric decays would amount to $N_\text{dec}(t_\text{dec})=N_\text{exc}(t_\text{exc})\cdot\left(1-e^{-\Gamma_\text{tot}t_\text{dec}}\right)\approx 2.7\cdot10^3$. About 25\% of the IC electrons can be expected to leave the target and, when accelerating the IC electrons onto an MCP detector, they could be detected with a relatively high efficiency of also about 50\%, leading to an expected total efficiency of $\sim12.5$\%.\\
Two types of background have to be considered in this experiment: (i) background in the form of electrons caused by the laser irradiation and (ii) low-energy electrons emitted during the $\alpha$ decay of $^{229}$Th and its daughter nuclei. Both types of background will be discussed in the following.\\
Photo-electrons will be emitted during laser irradiation. However, such electron emission terminates within femtoseconds after the end of a laser pulse. When considering time-of-flight effects, a pulse delay in the nanosecond range has to be taken into account. Still these processes are orders of magnitude faster than the isomeric decay, which allows to separate electrons emitted in the IC decay by means of the isomer's characteristic lifetime. Potentially slow electrons emitted in the surrounding of the target due to scattered light could be suppressed by a retarding field or removed by applying strong electric fields for a few nanoseconds triggered in coincidence with the laser pulses.\\
In case that the irradiated surface was heated up by the laser beam, thermionic electron emission would have to be considered as background, which typically continues on timescales of several 100~ns after the end of the laser pulse \cite{Ready}. While these timescales are still short enough to distinguish thermionic electron emission from the isomeric decay, it is also possible to quantitatively estimate this effect. The maximum surface temperature rise $\Delta T$ caused by a laser beam of absorbed irradiance $I$ after a laser pulse of duration $T_L$ (in case of vanishing optical penetration depth) is estimated by the equation \cite{Ready}

\begin{equation}
\Delta T=\frac{2I}{K}\cdot\left(\frac{\kappa T_L}{\pi}\right)^{1/2},
\label{surftemp}
\end{equation}
where $K$ denotes the thermal conductivity and $\kappa$ the thermal diffusivity of the bulk material. A single $T_L=5$~ns pulse of the considered laser system possesses an irradiance of $I=2\cdot10^5$~Wcm$^{-2}$, which is used as an upper bound for the actually absorbed intensity. If gold is used as the base material for the deposition of the $^{229}$Th layer, the thermal conductivity is $K\approx3.2$~Wcm$^{-1}$K$^{-1}$ and the thermal diffusivity $\kappa\approx1.3$~cm$^2$s$^{-1}$. This leads to a surface temperature rise of $\Delta T\approx 5.7$~K at the end of a laser pulse, which is negligible in terms of thermionic electron emission.\\
All high energy background effects originating from the radioactive source, can be efficiently filtered when placing the $^{229}$Th source in a magnetic bottle \cite{Yamakita}, as only electrons of below $\sim$100~eV kinetic energy will follow the magnetic field lines. The corresponding experimental detection scheme is shown in Fig.~\ref{Setup}. The $^{229}$Th coated substrate is placed in a magnetic bottle, consisting of a strong (about 0.5~T) permanent magnet and a magnetic coil, providing a weak magnetic field of about 1~mT. The MCP detector used for low-energy electron detection is placed on the opposite side of the coil. A hole is provided to allow for laser irradiation of the $^{229}$Th atoms. Note that the electrons will follow the magnetic field lines even if the coil region was curved. In this way there would be no direct line of sight between the detector and the source and only low-energy electrons have to be considered as background. Besides the isomeric decay, low-energy electrons are also emitted as a by-product during radioactive $\alpha$ decays. Typically, two low-energy electrons (from atomic shell re-organization following the $\alpha$ decay) are emitted per radioactive decay event \cite{Wandkowsky}. The intrinsic activity of $^{229}$Th of $\sim210$~Bq will soon increase by a factor of about $10$, due to daughter ingrowth, leading to expectedly about 4000 emitted electrons per second. However, as the detection time is 10~$\mu$s, there would be only about $4\cdot10^{-2}$ detected background electrons per pulse. This corresponds to a signal-to-background ratio of $S/B\approx7\cdot10^4$.\\
\begin{figure}
 \includegraphics[scale=1]{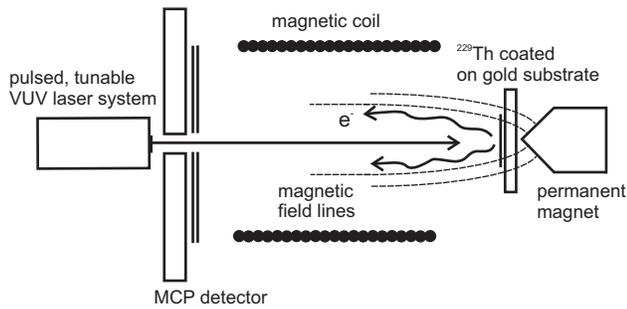}
\caption{Schematic drawing of the experimental concept, proposed to probe the direct laser excitation of the isomeric state in $^{229}$Th. See text for explanation.}
\label{Setup}
\end{figure}
Assuming that 100 laser pulses per scan step are acquired, the total time required for scanning the large energy range of 1~eV with the proposed laser of 10~GHz bandwidth would be about $2.4\cdot10^5$~s (corresponding to 2.7 days). In case of resonance, the expected amount of emitted IC electrons in 100 pulses would be $\sim2.7\cdot10^5$. In case of observing a signal, a comparative measurement using $^{232}$Th instead of $^{229}$Th could be performed in order to exclude surface effects as signal origin. In this way, the isomeric energy could be determined up to a fraction of the laser linewidth of 10~GHz (corresponding to a fraction of $4\cdot10^{-5}$~eV energy). However, as soon as the energy would be pinned down sufficiently precise, a laser system with a narrower bandwidth could be employed in order to further reduce the energy uncertainty.\\
\noindent In the following, the proposed detection of the isomeric excitation via the IC decay channel is compared to a different experimental approach, using $^{229}$Th-doped VUV-transparent crystals. The crystal-lattice approach, as proposed in Refs.~\cite{Peik1,Rellergert}, is closest to the presented idea, as also a high density of $^{229}$Th atoms is achieved. A VUV-transparent crystal (LiSrAlF$_6$) with a $^{229}$Th atomic density of $5.8\cdot10^{13}$~mm$^{-3}$ was already used in experiments \cite{Jeet}. The most important difference of this concept is that $^{229}$Th, when grown into the crystal, prefers the 4+ charge state. For this reason, and as the band gap of the crystal is large ($\sim10.7$~eV in case of LiSrAlF$_6$), non-radiative decay branches (like internal and bound-internal conversion) of $^{229\mathrm{m}}$Th are expected to be suppressed and the radiative decay is assumed to dominate \cite{Dessovic}. In this case $\alpha_\text{ic}=0$ should hold, leading to a long isomeric lifetime of up to $\sim10^4$~s.\\
An experiment is considered, in which a $^{229}$Th-doped crystal of 1~mm$^3$ in size (corresponding to $N_0\approx5.8\cdot10^{13}$ $^{229}$Th ions) is irradiated by laser light, using the same laser system as before. Assuming that a VUV spectrometer with a resolution of $\sim10^3$ is applied in order to reduce the background, the photon detection could be restricted to a wavelength interval of about 0.1~nm around the wavelength used for laser irradiation. It is emphasized that such spectrometer resolution is close to what is achievable with VUV prism spectrometers \cite{Moyssides}. A prism spectrometer, when combined with a high-efficient VUV optical system, could have the advantage of a higher light yield compared to the usual grating-based VUV spectrometer systems. The time intervals of laser irradiation and subsequent photon detection are chosen to about 100~s per scan step. As will be shown below, in this way a considerable number of isomeric decays should be detectable and the energy range of 1~eV could be scanned within $4.8\cdot10^6$~s (corresponding to 55.6 days).\\
Under these conditions, many individual laser pulses would be used for isomer excitation and the time-averaged laser spectral energy density has to be considered. This leads to

\begin{equation}
\rho^\omega=\frac{E_L R_L}{cA_L\cdot 2\pi\Delta\nu_L}\approx5.3\cdot10^{-18}\ \mathrm{Jm}^{-3}\mathrm{Hz}^{-1},
\end{equation}
with $E_L$ as the laser energy per pulse, $R_L$ the repetition rate, $A_L$ the irradiated area and $\Delta\nu_L$ the bandwidth of the laser light. According to Eq.~(\ref{laser1}), the number of excited nuclei after 100~s irradiation time will be $N_\text{exc}\approx3.1\cdot10^6$ and within further 100~s about $N_\text{dec}\approx3.1\cdot10^4$ isomeric decays would occur.\\
Generally, the lower limit for the background in the $^{229}$Th-doped crystal approach is given by Cherenkov radiation produced in the $\beta$ decays of short-lived daughter isotopes, unavoidably contained in the $^{229}$Th-doped crystals \cite{Stellmer}. Recently, the number of Cherenkov photons emitted along the entire decay chain per decay event of $^{229}$Th and per nanometer bandwidth (in the energy region of interest) was estimated to be $S_\lambda\approx 0.4$~nm$^{-1}$ \cite{Stellmer}. Beyond Cherenkov radiation, crystal fluorescence may
contribute to the background \cite{Stellmer2}. The number of $^{229}$Th atoms of $N_0\approx5.8\cdot10^{13}$ corresponds to a $^{229}$Th activity of 160~Bq. This in turn results in a number of detected Cherenkov photons after 100~s integration time of $640$ in the considered wavelength interval of $0.1$~nm. The signal-to-background ratio is thus inferred to be $S/B\approx48$.\\
While a corresponding experiment appears to be less favourable compared to the detection of IC electrons in terms of signal-to-background ratio and measurement time, it still seems to be realistic. However, it should be explicitly pointed out that the above detection scheme assumes a 100\% radiative decay of the isomer, which we consider as unlikely \cite{Wense3}. Any non-radiative decay, e.g., in the form of electronic bridge processes \cite{Karpeshin1}, might easily suppress the emission of light during the isomeric decay by several orders of magnitude, thereby preventing the isomer detection.\\
\noindent In conclusion, a new detection scheme for the direct laser excitation of $^{229\mathrm{m}}$Th was proposed that exploits the recently experimentally observed internal conversion decay channel of the $^{229}$Th isomeric state. The proposed experiment leads to an expected signal-to-background ratio of about $7\cdot10^4$ and a total measurement time of less than three days for scanning an energy range of 1~eV. This is by more than one order of magnitude more advantageous in both, signal-to-background ratio and measurement time, compared to an experiment which is based on the observation of a radiative isomeric decay channel. However, we see the most important advantage in the fact that it makes use of the internal conversion decay channel, which has recently been directly observed and can be securely expected to occur after laser excitation of the isomeric state for the presented experimental conditions.\\
We acknowledge discussions with T. Schumm, M. Laatiaoui, T. Udem, C. Düllmann, U. Morgner and J. Crespo López-Urrutia. This work was supported by the European Union's Horizon 2020 research and innovation programme under grant agreement 664732 "nuClock", by DFG Grants No. (Th956/3-1, Th956/3-2), and by the LMU department of Medical Physics via the Maier-Leibnitz Laboratory.

%\input acknowledgement.tex   % input acknowledgement

%\end{linenumbers}

\begin{thebibliography}{99}
\bibitem[1]{Evers} T.J. Bürvenich, J. Evers and C.H. Keitel, {\it Phys. Rev. Lett.} \textbf{96} 142501 (2006).
\bibitem[2]{Palffy} A. Pálffy, J. Evers and C.H. Keitel, {\it Phys. Rev. C} \textbf{77} 044602 (2008).
\bibitem[3]{Kroger_Reich} L.A. Kroger and C.W. Reich, {\it Nucl. Phys. A} \textbf{259} 29-60 (1976).
\bibitem[4]{Helmer_Reich2} R.G. Helmer and C.W. Reich, {\it Phys. Rev. C} \textbf{49} 1845-1859 (1994).
\bibitem[5]{Beck1} B.R. Beck, J.A. Becker, P. Beiersdorfer, G.V. Brown, K.J. Moody, J.B. Wilhelmy, F.S. Porter, C.A. Kilbourne and R.L. Kelley, {\it Phys. Rev. Lett.} \textbf{98} 142501 (2007).
\bibitem[6]{Beck2} B.R. Beck, C.Y. Wu, P. Beiersdorfer, G.V. Brown, J.A. Becker, K.J. Moody, J.B. Wilhelmy, F.S. Porter, C.A. Kilbourne and R.L. Kelley, {\it LLNL-PROC-415170} (2009).
\bibitem[7]{Tkalya1} E.V. Tkalya and V.O. Varlamov, {\it Phys. Scripta} \textbf{53} 296-299 (1996).
\bibitem[8]{Peik1} E. Peik and C. Tamm, {\it Eur. Phys. Lett.} \textbf{61} 181-186 (2003).
\bibitem[9]{Campbell2} C.J. Campbell, A.G. Radnaev, A. Kuzmich, V.A. Dzuba, V.V. Flambaum and A. Derevianko, {\it Phys. Rev. Lett.} \textbf{108} 120802 (2012).
\bibitem[10]{Rellergert} W.G. Rellergert, D. DeMille, R.R. Greco, M.P. Hehlen, J.R. Torgerson and E.R. Hudson, {\it Phys. Rev. Lett.} \textbf{104} 200802 (2010).
\bibitem[11]{Peik3} E. Peik and M. Okhapkin, {\it C.R. Physique} \textbf{16} 516-523 (2015).
\bibitem[12]{Ruchowska} E. Ruchowska, W.A. Plociennik and J. Zylicz {\it et al.}, {\it Phys. Rev. C} \textbf{73} 044326 (2006).
\bibitem[13]{Tkalya4} E.V. Tkalya, C. Schneider, J. Jeet and E.R. Hudson, {\it Phys. Rev. C} \textbf{92} 054324 (2015).
\bibitem[14]{Minkov} N. Minkov and A. Pálffy, {\it Phys. Rev. Lett.} \textbf{118} 212501 (2017).
\bibitem[15]{Seiferle4} B. Seiferle, L. von der Wense and P.G. Thirolf, Eur. Phys. J. A \textbf{53} (2017) 108.
\bibitem[16]{Kazakov} G.A. Kazakov, V. Schauer, J. Schwestka, S.P. Stellmer, J.H. Sterba, A. Fleischmann, L. Gastaldo, A. Pabinger, C. Enss and T. Schumm, {\it Nucl. Instrum. Methods A} \textbf{735} 229-239 (2014).
\bibitem[17]{Yamaguchi} A. Yamaguchi, M. Kolbe, H. Kaser, T. Reichel, A. Gottwald and E. Peik, {\it New J. Phys.} \textbf{17} 053053 (2015).
\bibitem[18]{Jeet} J. Jeet, C. Schneider, S.T. Sullivan, W.G. Rellergert, S. Mirzadeh, A. Cassanho, H.P. Jenssen, E.V. Tkalya and E.R. Hudson, {\it Phys. Rev. Lett.} \textbf{114} 253001 (2015).
\bibitem[19]{Wense1} L. von der Wense, B. Seiferle, M. Laatiaoui, J.B. Neumayr, H.J. Maier, H.F. Wirth, C. Mokry, J. Runke, K. Eberhardt, C.E. Düllmann, N.G. Trautmann and P.G. Thirolf, {\it Nature} \textbf{533} 47-51 (2016).
\bibitem[20]{Seiferle3} B. Seiferle, L. von der Wense and P.G. Thirolf, {\it Phys. Rev. Lett.} \textbf{118} 042501 (2017).
\bibitem[21]{Salvat} F. Salvat and J. Parellada, {\it Nucl. Instrum. Meth.} B1 70-84 (1984).
\bibitem[22]{Loudon} R. Loudon, {\it The quantum theory of light}, Oxford University Press (2003).
\bibitem[23]{Hilborn} R.C. Hilborn, Am. J. Phys. \textbf{50} (1982) 982-986.
\bibitem[24]{Wense3} L. von der Wense, Ph.D. thesis, Ludwig-Maximilians-Universität München, Germany (2016). Online available at: {\it https://edoc.ub.uni-muenchen.de/20492/7/ Wense\_Lars\_von\_der.pdf}
\bibitem[25]{Trautmann} N. Trautmann, {\it J. Alloy. Compd.} \textbf{213-2014} 28-32 (1994).
\bibitem[26]{Strizhov} V.F. Strizhov and E.V. Tkalya, {\it Sov. Phys. JETP} \textbf{72} 387-390 (1991).
\bibitem[27]{Karpeshin1} F.F. Karpeshin and M.B. Trzhaskovskaya, {\it Phys. Rev C} \textbf{76}, 054313  (2007).
\bibitem[28]{Hanna} S.J. Hanna, P. Campuzano-Jost, E.A. Simpson, D.B. Robb, I. Burak, M.W. Blades, J.W. Hepburn and A.K. Bertram, {\it Int. J. Mass Spectrom.} \textbf{279} 134-146 (2009).
\bibitem[29]{Dai} S.B. Dai, M. Chen, S.J. Zhang, Z.M. Wang, F.F. Zhang, F. Yang, Z.C. Wang, N. Zong, L.J. Liu and X.Y. Wang, {\it Laser Phys. Lett.} \textbf{13} 035401 (2016).
\bibitem[30]{Hilbig} R. Hilbig and R. Wallenstein, {\it Applied Optics} \textbf{21} 913-917 (1982).
\bibitem[31]{Zimmermann} K. Zimmermann, {\it Ph.D. Thesis} Univ. Hannover, Germany (2010).
\bibitem[32]{Gill} C.G. Gill, T.M. Allen, J.E. Anderson, T.N. Taylor, P.B. Kelly and N.S. Nogar, {\it Applied Optics} \textbf{35} 2069-2082 (1996).
\bibitem[33]{Riviere} J.C. Rivière, {\it Proc. Phys. Soc.} \textbf{80} 124-129 (1962).
\bibitem[34]{Seah} M.P. Seah and W.A. Dench, {\it Surface and Interface Analysis} \textbf{1} 2-11 (1979).
\bibitem[35]{Riviere2} J.C. Rivière, {\it Brit. J. Appl. Phys.} \textbf{16} 1507-1511 (1965).
\bibitem[36]{Hollenshead} J. Hollenshead and L. Klebanoff, {\it J. Vac. Sci. Technol. B} \textbf{24} 64-82 (2006).
\bibitem[37]{Canfield} L.R. Canfield, G. Hass and W.R. Hunter, {\it Le Journal de Physique} \textbf{25} 124-129 (1964).
\bibitem[38]{Ready} J.F. Ready, {\it Academic Press} (1971).
\bibitem[39]{Yamakita} Y. Yamakita, H. Tanaka, R. Maruyama, H. Yamakado, F. Misaizu and K. Ohno, {\it Rev. Sci. Instrum.} \textbf{71} 3042-3049 (2000).
\bibitem[40]{Wandkowsky} N. Wandkowsky, G. Drexlin, F.M. Fränkle, F. Glück, S. Groh and S. Mertens, {\it New J. Phys.} \textbf{15} 083040 (2013).
\bibitem[41]{Dessovic} P. Dessovic, P. Mohn, R.A. Jackson, M. Schreitl, G. Kazakov and T. Schumm, {\it J. Phys. Condens. Matter} \textbf{26} 105402 (2014).
\bibitem[42]{Moyssides} P.G. Moyssides, S. Maltezos and E. Fokitis, {\it J. Mod. Optics} \textbf{47} 1693-1706 (2000).
\bibitem[43]{Stellmer} S. Stellmer, M. Schreitl, G.A. Kazakov, J.H. Sterba and T. Schumm, {\it Phys. Rev. C} \textbf{94} 014302 (2016).
\bibitem[44]{Stellmer2} S. Stellmer, M. Schreitl and T. Schumm, {\it Sci. Rep.} \textbf{5} 15580 (2015).


\end{thebibliography}
\end{document}